\documentclass[10pt]{article}
\usepackage{multicol}
\usepackage{graphicx}
\usepackage{amsmath}
\usepackage[a4paper]{geometry}
\usepackage{verbatim}
\usepackage{hyperref}
\usepackage[usenames,dvipsnames]{color}
\usepackage{isomath}

\definecolor{darkblue}{RGB}{0,0,196}

\begin{document}
\begin{center}

{\Large\bf Fragmentation of Nuclear Remnants in Electron-Nucleus
Collisions at High Energy as a Nonextensive Process}\vspace{0.5cm}

Ting-Ting~Duan$^{1,}${\footnote{202312602001@email.sxu.edu.cn}},
Sahanaa~B$\rm\ddot{u}$riechin$^{1,}${\footnote{202201101236@email.sxu.edu.cn}},
Hai-Ling~Lao$^{2,}${\footnote{hailinglao@163.com;
hailinglao@pku.edu.cn}},
Fu-Hu~Liu$^{1,}${\footnote{Correspondence: fuhuliu@163.com;
fuhuliu@sxu.edu.cn}},
Khusniddin~K.~Olimov$^{3,4,}${\footnote{Correspondence:
khkolimov@gmail.com; kh.olimov@uzsci.net}}

{\small\it $^1$Institute of Theoretical Physics, State Key
Laboratory of Quantum Optics Technologies and Devices \& \\
Collaborative Innovation Center of Extreme Optics, Shanxi
University, Taiyuan 030006, China

$^2$Department of Science Teaching, Beijing Vocational College of
Agriculture, Beijing 102442, China

$^3$Laboratory of High Energy Physics, Physical-Technical
Institute of Uzbekistan Academy of Sciences, \\ Chingiz Aytmatov
Str. 2b, Tashkent 100084, Uzbekistan

$^4$Department of Natural Sciences, National University of Science
and Technology MISIS (NUST MISIS), \\ Almalyk Branch, Almalyk
110105, Uzbekistan}

\end{center}


\vspace{0.5cm}

\noindent {\bf Abstract:} Utilizing a partitioning method based on
equal (or unequal) probabilities---without incorporating the
alpha-cluster ($\alpha$-cluster) model---allows for the derivation
of diverse topological configurations of nuclear fragments
resulting from fragmentation. Subsequently, we predict the
multiplicity distribution of nuclear fragments for specific
excited nuclei, such as $^9$Be$^*$, $^{12}$C$^*$, and
$^{16}$O$^*$, which can be formed as nuclear remnants in
electron-nucleus ($eA$) collisions at high energy.
Based on the $\alpha$-cluster model, an
$\alpha$-cluster structure may result in deviations in the
multiplicity distributions of nuclear fragments with charge $Z=2$,
compared to those predicted by the partitioning methods.
Furthermore, in the framework of Tsallis statistics, the
nonextensive generalized temperature, entropy index, and
$q$-entropy are obtained from the multiplicity distribution of
nuclear fragments with given charge number. Our work shows that
fragmentation of nuclear remnants in electron-nucleus collisions
at high energy is a nonextensive process.
\\
\\
{\bf PACS Number(s):} 21.60.Gx, 21.60.-n, 21.60.Ka
\\
\\


\parindent=15pt


{\section{Introduction}}

High-energy nuclear collisions represent a significant area of
research within modern physics~\cite{a,a1,b,b1,b2}. In these
interactions, numerous particles are predominantly generated
within participant regions while multiple fragments are primarily
emitted from spectator regions when available. A
series of recent research achievements focusing on the field of
nuclear fragmentation---spanning model construction, experimental
planning, and mechanism analysis---comprehensively demonstrate the
vigorous development trend of this field, providing multiple
perspectives for a deeper understanding of nuclear structure and
nuclear reaction dynamics~\cite{b3,b4,b5,b6,b7,b8,b9}.

Relevant research content includes, but is not
limited to, the phenomenon of nuclear fragmentation in the
intermediate energy region, the influence of nuclear collective
excitation modes on nuclear fragmentation, the competition between
nuclear fragmentation and electromagnetic fragmentation, how
nuclear micro-excited states affect the production of macroscopic
nuclear fragments, and the application of Monte Carlo methods in
nuclear fragmentation studies~\cite{b3,b4,b5,b6,b7,b8,b9}.
These studies complement and advance one another,
not only propelling the field of nuclear fragmentation forward but
also laying a solid foundation for progress in related areas such
as nuclear physics and nuclear technology applications.

During multi-fragment emission processes, it is anticipated that
spectators will form an excited nucleus. This excited nucleus
subsequently undergoes fragmentation into various
components~\cite{c,d,e,f}. Such fragmentation reveals rich
internal structures within the nucleus where protons and neutrons
can combine appropriately to create intermediate
configurations~\cite{1,2,3,4}. When two protons and two neutrons
coalesce into such an intermediate structure, it is referred to as
an alpha-cluster ($\alpha$-cluster) structure~\cite{5,6,7,8}. It
is possible that two or more $\alpha$-cluster structures are
existent in a heavy nucleus~\cite{9,10,11,12,13}. According to the
$\alpha$-cluster model~\cite{14,15,16,17,18}, the $\alpha$-cluster
structure should have much higher probability than other
intermediate structures.

Different configurations of nuclear fragments can be measured in
the fragmentation of excited nuclei. A multi-$\alpha$
configuration is one such arrangement that arises from statistical
or stochastic fragmentation processes. Several experimental
results concerning multi-$\alpha$ configurations in $^{16}$O
fragmentation at high energies have been reported in the
literature~\cite{c,19,20,21,22}. According to the $\alpha$-cluster
model~\cite{14,15,16,17,18}, further evidence and a higher
probability for the presence of $\alpha$-cluster structures are
anticipated. If the occurrence of multi-$\alpha$ configurations is
significantly more probable than what would be expected based on
partitioning methods derived from stochastic processes, it may
indeed reflect the underlying $\alpha$-cluster structure of the
excited nucleus.

Due to challenges in excluding the influence of partitioning
probabilities across various configurations, there remains limited
experimental evidence supporting $\alpha$-cluster structures; this
evidence is insufficient for a comprehensive validation of the
$\alpha$-cluster model~\cite{14,15,16,17,18}. Consequently, there
is a pressing need for more systematic experimental
investigations. This necessity motivates researchers to measure
fragmentation products originating from excited nuclei. It is
anticipated that diverse types of excited nuclei (nuclear
remnants) will be formed through nuclear reactions induced by
electrons at the forthcoming Electron-Ion Collider
(EIC)~\cite{23}. The EIC presents an exceptional opportunity for
researchers to systematically explore $\alpha$-cluster structures
and thoroughly validate the $\alpha$-cluster model using
fragmentation products from excited nuclei.

Furthermore, nuclear fragmentation is a non-thermal equilibrium
process, rendering Boltzmann-Gibbs statistics inapplicable.
Instead, one may adopt two-, three-, or multi-component
distributions, where each component represents a state of local
equilibrium that can be described within the framework of
Boltzmann-Gibbs statistics. Consequently, a multi-temperature
pattern, or temperature fluctuations, can be observed in nuclear
fragmentation processes. Empirically, such multi-component
distributions can be effectively fitted using Tsallis statistics,
which indicates that nuclear fragmentation is inherently a
nonextensive process. Therefore, nonextensive parameters derived
from Tsallis statistics, including generalized temperature,
entropy index, and $q$-entropy, can be employed to characterize
and analyze nuclear fragmentation phenomena.

In this work, both equal and unequal probability partitioning
methods are employed to derive various configurations of nuclear
fragments from excited states such as $^9$Be$^*$, $^{12}$C$^*$,
and $^{16}$O$^*$, which are expected to result from
electron-nucleus ($eA$) collisions at the EIC~\cite{23}. The
multiplicity distributions for all fragments as well as those with
charge $Z$ will subsequently be obtained. In particular, the
probability of multi-$\alpha$ or multi-He configuration or channel
can be obtained, which may serve as the baseline for judging about
the $\alpha$-cluster structure. In addition, in the framework of
Tsallis statistics, the nonextensive generalized temperature,
entropy index, and $q$-entropy are obtained from the multiplicity
distribution of nuclear fragments with given charge number.

The remainder of this paper is structured as follows. Various
configurations of nuclear fragments are described in Section 2.
Multiplicity distributions of nuclear fragments are presented in
Section 3. In Section 4, we show nonextensive parameters from
multiplicity distributions of nuclear fragments. Finally, we
give summary and conclusion of this work in
Section 5.
\\

{\section{Various configurations of nuclear fragments}}

In the context of multi-fragment emission during $eA$ collisions,
various fragmentation properties warrant special attention. For
instance, understanding the types of fragments is crucial for
elucidating the mechanisms underlying nuclear fragmentation;
however, determining the number of neutrons in an isotope presents
a complex challenge. Our previous research has demonstrated that
the isotopic production cross section follows an Erlang
distribution~\cite{23a}. When different isotopes with a given
charge number are not distinguished from one another, the analysis
becomes significantly more straightforward.

At the EIC, as nuclear remnants, excited nuclei formed in $eA$
collisions can fragment into diverse topological configurations.
This allows for an investigation into the internal structure of
these excited nuclei. During fragmentation, both proton and
neutron numbers are conserved. In experimental settings, it is
possible to measure either the charge or proton count of a
fragment. This capability facilitates our examination of
multiplicity distributions among fragments with varying charges.
Furthermore, we may delve deeper into discussing the fundamental
physical reasons behind these multiplicity distributions observed
in nuclear fragments. Notably, factors such as $\alpha$-cluster
structures and liquid-gas phase transitions could influence these
experimentally measured multiplicity distributions.

As examples, we now consider three types of $eA$ collisions at the
EIC:
\begin{align}
e+ ^{10}{\rm Be}\longrightarrow \bigg\{\begin{aligned} &(e+n)+^9{\rm Be}^*\\
&(e+p)+^9{\rm Li}^*, \ \ ^9{\rm Li}^* \longrightarrow 2n+^7{\rm
Li}^* \ \ {\rm or} \ \ 3n+^6{\rm Li}^*,
\end{aligned}
\end{align}
\begin{align}
e+ ^{13}{\rm C}\longrightarrow \bigg\{\begin{aligned} &(e+n)+^{12}{\rm C}^*\\
&(e+p)+^{12}{\rm B}^*, \ \ ^{12}{\rm B}^* \longrightarrow
n+^{11}{\rm B}^* \ \ {\rm or} \ \ 2n+^{10}{\rm B}^*,
\end{aligned}
\end{align}
and
\begin{align}
e+ ^{17}{\rm O}\longrightarrow \bigg\{\begin{aligned} &(e+n)+^{16}{\rm O}^*\\
&(e+p)+^{16}{\rm N}^*, \ \ ^{16}{\rm N}^* \longrightarrow
n+^{15}{\rm N}^* \ \ {\rm or} \ \ 2n+^{14}{\rm N}^*.
\end{aligned}
\end{align}
Then, the excited $^9$Be, $^{12}$C, and $^{16}$O nuclei can be
obtained and analyzed. Other excited nuclei are not the focus of
the present work due to the fact that they do not have an
advantage in the study of $\alpha$-cluster structure.

It is important to note that the process $e+n$ or $e+p$ occurring
in $eA$ is not electron-induced neutron/proton knock-out reaction,
which typically manifests at beam energies of hundreds of
MeV~\cite{23b,23c}. Instead, this represents multi-particle
production process that occurs at beam energies on the order of
hundreds of GeV, for which the EIC is specifically designed to
achieve~\cite{23d}, with a center-of-mass energy range between 20
and 100 GeV. In the incident nucleus $A$, alongside the
participant nucleon, there exist spectator nucleons---the
remaining constituents---which will form an excited nucleus
characterized by energy levels significantly higher than those
attainable through MeV collisions.

The excited nuclei subsequently decay into various nuclear
fragments. The correlations between momentum and scattering angle
for evaporated neutrons and protons have been extensively studied
using the BeAGLE (Benchmark eA Generator for LEptoproduction)
model~\cite{23e}, particularly in high-energy lepton-nucleus
collisions. The excited nuclei produced in $eA$ collisions at the
EIC exhibit significantly higher excitation levels compared to
those generated in electron-induced neutron/proton knock-out
reactions conducted with fixed targets at low and medium energies.
Due to substantial excitation leading to large internal momenta,
both decay protons and other nuclear fragments correspond to
sufficiently large polar angles that fall within the estimated
pseudorapidity acceptance region designated for the currently
proposed EIC detector~\cite{23,23f}.

Considering the Fermi momentum of a nucleon within the nucleus,
which is approximately 0.25 GeV$/c$, and the momentum per nucleon
of the incident nucleus being 10 GeV$/c$, decay protons and other
fragments are expected to be emitted within a forward cone
characterized by a polar angle $\theta_0=25$ mrad. This
corresponds to a pseudorapidity of
$\eta=-\ln\tan(\theta_0/2)=4.38$. Furthermore, recoil protons can
be distinguished from decay protons since recoil protons
participate in multi-scattering processes which result in much
larger scattering angles. It is assumed that this emission will
span a wide range from nearly 0 (corresponding to $\eta=\infty$)
up to approximately $10\theta_0$ (which corresponds to
$\eta=2.07$). Indeed, it cannot be excluded that some recoil
protons may have very small scattering angles, leading to
exceptionally large pseudorapidities due to the influence of
leading nucleons.

If the two types of protons are assumed to emit isotropically in
their respective rest frames, they approximately follow Gaussian
$\eta$ distributions with a common standard deviation
($\sigma_{\eta}\approx0.91$). The decay protons predominantly
distribute within the range of
$4.38<\eta<4.38+4\sigma_{\eta}=8.02$, while the recoil protons
primarily occupy the range of
$2.07<\eta<2.07+4\sigma_{\eta}=5.71$. It is evident that there
exists some overlap between decay and recoil protons in the
forward cone. Although most recoil protons may have emission
angles exceeding 25 mrad due to multiple scattering processes, we
cannot entirely dismiss the possibility of them appearing within
the forward cone. To more effectively distinguish between decay
and recoil protons, one could study their energies; generally, the
energy of a decay proton is nearly equal to that of an incident
nucleus per nucleon, whereas the energy of a recoil proton should
be lower.

While it is challenging to precisely separate decay from recoil
protons in the forward cone, such distinction is not essential for
this study. In fact, among the three types of $eA$ collisions
considered in previously mentioned reactions (1)--(3), our
selected samples should ideally consist solely of those with only
recoil neutrons; thus, any contributions from recoil protons must
be excluded from our analysis. In rare instances where mixed
events occur involving recoiling protons within the forward cone,
these can introduce minor measurement errors. In cases where
recoiling protons do appear in this region, misidentified events
would include both these recoils and fragmentations from $^9{\rm
Li}^*$, $^{12}{\rm B}^*$, and $^{16}{\rm N}^*$
respectively---these should be excluded from expected events
associated with fragmentations originating from $^9{\rm Be}^*$,
$^{12}{\rm C}^*$, and $^{16}{\rm O}^*$. Although there are clear
differences in energy between decay and recoil protons, attempting
to separate them experimentally may incur additional costs.

In our studies concerning nuclear fragments, our primary focus
lies on counting electric charges rather than mass, momentum,
energy, etc. The resolutions of detectors regarding secondary
quantities do not impact our analysis; however, it is crucial that
detector resolution for charge number remains high---approximately
$\sim2\%$, which is generally achievable. During experiments, it
is essential to select relevant decay events where the total
charge number of various nuclear fragments precisely matches that
of the incident nucleus $A$. In some events, due to distribution
fluctuations, individual nuclear fragment has probability to emit
with a very small polar angle, which is mixed with the beam and
cannot be captured by the detector~\cite{23,23d,23f}. Naturally,
these events should be removed from the analysis.

In addition to $N_F$ representing the multiplicity of all nuclear
fragments, let $N_Z$ be the multiplicity of the fragments with
charge $Z$. In an equal probability partitioning method, in which
the $\alpha$-cluster model does not enter, the frequency of
configuration $\{N_Z(Z)\}$, or the weight of partition
$\{N_Z(Z)\}$, is considered to be the same which results in the
same probability $f_1$. Various topological configurations of
nuclear fragments in excited nuclear fragmentation can be obtained
by the treatment of exhaustive enumeration.

In an unequal probability partitioning method~\cite{24,25}, in
which the $\alpha$-cluster model does not enter either, the
frequency of configuration $\{N_Z(Z)\}$, or the weight of
partition $\{N_Z(Z)\}$, is considered to be the number of exchange
\begin{align}
M_2=\frac{Q!}{\prod\limits_Z N_Z(Z)!Z^{N_Z}},
\end{align}
where $Q$ is the charge number of the excited nucleus, $Q!$ and
$N_Z(Z)!$ represent factorial operations, and $M_2$ is the Cauchy
number in the number theory. The normalization of $M_2$ is
\begin{align}
\sum\limits_{\{N_Z(Z)\}} M_2=Q!.
\end{align}
The probability of configuration $\{N_Z(Z)\}$ is
\begin{align}
f_2=\frac{M_2}{\sum\limits_{\{N_Z(Z)\}} M_2}=
\frac{1}{\prod\limits_Z N_Z(Z)!Z^{N_Z}}.
\end{align}

The equal and unequal probability partitioning methods present
distinct perspectives in the realm of physics. The equal
probability partitioning method is grounded in the principle of
equal probability, a fundamental assumption in statistical
physics. This principle asserts that when a system is at
equilibrium, provided there are no additional constraints beyond
energy, volume, and particle number, the likelihood of the system
occupying each microscopic state remains uniform. Conversely, the
unequal probability partitioning method relies on the principle of
unequal probability; this acknowledges that within a sampling
survey, the chance of selecting any individual from a population
may vary due to the interchangeability of identical particles.

Prior to implementing the partitioning methods, it is essential to
highlight other applications that demonstrate their validity and
rationality. In previous studies~\cite{25a,25b,25c,25d,25e}, these
methods were employed to investigate excited nuclear fragmentation
during nucleus-nucleus collisions at intermediate and high
energies. Conditional moments and their normalized forms across
various orders were introduced~\cite{25a,25b} for examining
critical behavior~\cite{25f,25g}. It was observed that
correlations and distributions derived from conditional moments of
nuclear fragments~\cite{25c,25d,25e} obtained through the
partitioning technique~\cite{24, 25} align well with experimental
data concerning excited nuclear fragmentation resulting from
diffractive excitation (nuclear reaction) as well as
electromagnetic dissociation~\cite{25h,25i}.

Using the equal (unequal) probability partitioning method, various
topological configurations of nuclear fragments in excited $^9$Be,
$^{12}$C, and $^{16}$O fragmentation are listed in Tables 1, 2,
and 3, respectively, in which each configuration has an equal
(unequal) probability $f_1$ ($f_2$). The multiplicity, $N_F$, of
all fragments and the multiplicity, $N_Z$, of the fragments with
given charge $Z$ in a defined configuration are shown separately.

In the equal probability partitioning method, the numbers of
configurations, or fragmentation channels, in fragmentation of
excited $^9$Be, $^{12}$C, and $^{16}$O nuclei are 5, 11, and 22,
respectively. The fragment Be is artificially assumed by default
with 50\% probability to be the most unstable $^8$Be and in 50\%
of the cases to be (relative) stable isotope of Be. $^8$Be can
decay into 2He, which is listed in brackets with fractions in the
tables. In the unequal probability partitioning method, the
numbers of exchanges in excited $^9$Be, $^{12}$C, and $^{16}$O
fragmentation are 24, 720, and 40320, respectively. The fragment
Be is assumed by default with a given chance ($\{M_2(2{\rm
He})/[M_2(2{\rm He})+M_2({\rm Be})]=1/3\}$) to be $^8$Be which is
unstable and can decay into 2He with given fractions.
\\

\begin{table*}[htbp]
{\small Table 1: The multiplicity, $N_F$, of all fragments and the
multiplicity, $N_Z$, of the fragments with charge $Z$ in various
configurations in excited $^9$Be fragmentation, where only the
charge conservation is considered in the fragmentation. In the
equal probability partitioning method, the fragment Be is
defaulted with 50\% probability to be $^8$Be, and in the unequal
probability partitioning method, the fragment Be is defaulted with
a given chance $\{M_2(2{\rm He})/[M_2(2{\rm He})+M_2({\rm
Be})]=1/3\}$ to be $^8$Be, where $^8$Be is unstable and can decay
into 2He, which is listed in the bracket, and causes $N_F$ to
$N_F+1$ and $N_{Z=2}$ to $N_{Z=2}+2$. Here, the changeable $N_F$
and $N_{Z=2}$ are shown in the table by $+1$ and $+2$,
respectively. The probabilities $f_1$ and $f_2$ of each channel
obtained by the equal and unequal partitioning methods are listed
respectively.} \vspace{-5mm} {\small
\begin{center}
\begin{tabular}{cccccccc}\\
\hline\hline
$N_F$ & $N_{Z=1}$ & $N_{Z=2}$ & $N_{Z=3}$ & $N_{Z=4}$ & Configuration & $f_1$ (1/5) & $f_2$ (1/24)\\
\hline
4 & 4 &   &   &   & 4H    & 1   & 1\\
\hline
3 & 2 & 1 &   &   & 2H+He & 1   & 6\\
\hline
2 & 1 &   & 1 &   & H+Li  & 1   & 8\\
2 &   & 2 &   &   & 2He   & 1   & 3\\
\hline
1 &   &   &   & 1 & [Be   & 0.5 & 4\\
1+1&  &+2 &   &   & (2He)]& 0.5 & 2\\
\hline
\end{tabular}%
\end{center}}
\end{table*}

\begin{table*}[htb!]
{\small Table 2: The multiplicity $N_F$ of all fragments and the
multiplicity $N_Z$ of the fragments with charge $Z$ in various
configurations in excited $^{12}$C fragmentation, where only the
charge conservation is considered in the fragmentation. In the
equal probability partitioning method, the fragment Be is
defaulted with 50\% probability to be $^8$Be, and in the unequal
probability partitioning method, the fragment Be is defaulted with
a given chance (1/3) to be $^8$Be, where $^8$Be is unstable and
can decay into 2He, which is listed in the bracket, and causes
$N_F$ to $N_F+1$ and $N_{Z=2}$ to $N_{Z=2}+2$. The probabilities
$f_1$ and $f_2$ of each channel obtained by the equal and unequal
partitioning methods are listed respectively.} \vspace{-5mm}
{\small
\begin{center}
\begin{tabular}{cccccccccc}\\
\hline\hline
$N_F$ & $N_{Z=1}$ & $N_{Z=2}$ & $N_{Z=3}$ & $N_{Z=4}$ & $N_{Z=5}$ & $N_{Z=6}$ & Configuration & $f_1$ (1/11) & $f_2$ (1/720)\\
\hline
6 & 6 &   &   &   &   &   & 6H       & 1   & 1\\
\hline
5 & 4 & 1 &   &   &   &   & 4H+He    & 1   & 15\\
\hline
4 & 3 &   & 1 &   &   &   & 3H+Li    & 1   & 40\\
4 & 2 & 2 &   &   &   &   & 2H+2He   & 1   & 45\\
\hline
3 & 2 &   &   & 1 &   &   & [2H+Be   & 0.5 & 60\\
3+1&2 &+2 &   &   &   &   & 2H+(2He)]& 0.5 & 30\\
3 & 1 & 1 & 1 &   &   &   & H+He+Li  & 1   & 120\\
3 &   & 3 &   &   &   &   & 3He      & 1   & 15\\
\hline
2 & 1 &   &   &   & 1 &   & H+B      & 1   & 144\\
2 &   & 1 &   & 1 &   &   & [He+Be   & 0.5 & 60\\
2+1&  &1+2&   &   &   &   & He+(2He)]& 0.5 & 30\\
2 &   &   & 2 &   &   &   & 2Li      & 1   & 40\\
\hline
1 &   &   &   &   &   & 1 & C        & 1   & 120\\
\hline
\end{tabular}%
\end{center}}
\end{table*}

\begin{table*}[htb!]
{\small Table 3: The multiplicity $N_F$ of all fragments and the
multiplicity $N_Z$ of the fragments with charge $Z$ in various
configurations in excited $^{16}$O fragmentation, where only the
charge conservation is considered in the fragmentation. In the
equal probability partitioning method, the fragment Be is
defaulted with 50\% probability to be $^8$Be, and in the unequal
probability partitioning method, the fragment Be is defaulted with
a given chance (1/3) to be $^8$Be, where $^8$Be is unstable and
can decay into 2He, which is listed in the bracket, and causes
$N_F$ to $N_F+1$ and $N_{Z=2}$ to $N_{Z=2}+2$. The probabilities
$f_1$ and $f_2$ of each channel obtained by the equal and unequal
partitioning methods are listed respectively.} \vspace{-10mm}
{\small
\begin{center}
\begin{tabular}{cccccccccccc}\\
\hline\hline
$N_F$ & $N_{Z=1}$ & $N_{Z=2}$ & $N_{Z=3}$ & $N_{Z=4}$ & $N_{Z=5}$ & $N_{Z=6}$ & $N_{Z=7}$ & $N_{Z=8}$ & Configuration & $f_1$ (1/22) & $f_2$ (1/40320)\\
\hline
8 & 8 &   &   &   &   &   &   &   & 8H     & 1 & 1\\
\hline
7 & 6 & 1 &   &   &   &   &   &   & 6H+He  & 1 & 28\\
\hline
6 & 5 &   & 1 &   &   &   &   &   & 5H+Li  & 1 & 112\\
6 & 4 & 2 &   &   &   &   &   &   & 4H+2He & 1 & 210\\
\hline
5 & 4 &   &   & 1 &   &   &   &   & [4H+Be    & 0.5 & 280\\
5+1&4 &+2 &   &   &   &   &   &   & 4H+(2He)] & 0.5 & 140\\
5 & 3 & 1 & 1 &   &   &   &   &   & 3H+He+Li  & 1   & 1120\\
5 & 2 & 3 &   &   &   &   &   &   & 2H+3He    & 1   & 420\\
\hline
4 & 3 &   &   &   & 1 &   &   &   & 3H+B         & 1   & 1344\\
4 & 2 & 1 &   & 1 &   &   &   &   & [2H+He+Be    & 0.5 & 1680\\
4+1&2 &1+2&   &   &   &   &   &   & 2H+He+(2He)] & 0.5 & 840\\
4 & 2 &   & 2 &   &   &   &   &   & 2H+2Li       & 1   & 1120\\
4 & 1 & 2 & 1 &   &   &   &   &   & H+2He+Li     & 1   & 1680\\
4 &   & 4 &   &   &   &   &   &   & 4He          & 1   & 105\\
\hline
3 & 2 &   &   &   &   & 1 &   &   & 2H+C        & 1   & 3360\\
3 & 1 & 1 &   &   & 1 &   &   &   & H+He+B      & 1   & 4032\\
3 & 1 &   & 1 & 1 &   &   &   &   & [H+Li+Be    & 0.5 & 2240\\
3+1&1 &+2 & 1 &   &   &   &   &   & H+(2He)+Li] & 0.5 & 1120\\
3 &   & 2 &   & 1 &   &   &   &   & [2He+Be     & 0.5 & 840\\
3 &   &2+2&   &   &   &   &   &   & 2He+(2He)]  & 0.5 & 420\\
3 &   & 1 & 2 &   &   &   &   &   & He+2Li      & 1   & 1120\\
\hline
2 & 1 &   &   &   &   &   & 1 &   & H+N          & 1    & 5760\\
2 &   & 1 &   &   &   & 1 &   &   & He+C         & 1    & 3360\\
2 &   &   & 1 &   & 1 &   &   &   & Li+B         & 1    & 2688\\
2 &   &   &   & 2 &   &   &   &   & [2Be         & 0.25 & 504\\
2+1&  &+2 &   & 1 &   &   &   &   & (2He)+Be     & 0.5  & 504\\
2+1+1& &+2+2& &   &   &   &   &   & (2He)+(2He)] & 0.25 & 252\\
\hline
1 &   &   &   &   &   &   &   & 1 & O & 1 & 5040\\
\hline
\end{tabular}%
\end{center}}
\end{table*}

\begin{table*}[htbp]
{\small Table 4: The multiplicity distribution, $dn/dN_F$, of all
fragments and the multiplicity distribution, $dn/dN_Z$, of the
fragments with charge $Z$ in excited $^9$Be fragmentation in the
equal (unequal) probability partitioning method, where the
normalization is 5 (24), which is the number of total
configurations (exchanges). In the table, $N_x$ denotes $N_F$ or
$N_Z$.} \vspace{-5mm} {\small
\begin{center}
\begin{tabular}{cccccc}\\
\hline\hline
$dn/dN_x$ & $N_x=0$ & $N_x=1$ & $N_x=2$ & $N_x=3$ & $N_x=4$\\
\hline
$dn/dN_F$     &  0 (0)    & 0.5 (4) & 2.5 (13) & 1 (6) & 1 (1)\\
$dn/dN_{Z=1}$ &  2 (9)    & 1 (8)   & 1 (6)    & 0 (0) & 1 (1)\\
$dn/dN_{Z=2}$ &  2.5 (13) & 1 (6)   & 1.5 (5)  & 0 (0) & 0 (0)\\
$dn/dN_{Z=3}$ &  4 (16)   & 1 (8)   & 0 (0)    & 0 (0) & 0 (0)\\
$dn/dN_{Z=4}$ &  4.5 (20) & 0.5 (4) & 0 (0)    & 0 (0) & 0 (0)\\
\hline
\end{tabular}%
\end{center}}
\end{table*}

\begin{table*}[htb!]
{\small Table 5: The multiplicity distribution $dn/dN_F$ of all
fragments and the multiplicity distribution $dn/dN_Z$ of the
fragments with charge $Z$ in excited $^{12}$C fragmentation in the
equal (unequal) probability partitioning method, where the
normalization is 11 (720), which is the number of total
configurations (exchanges).} \vspace{-5mm} {\small
\begin{center}
\begin{tabular}{cccccccc}\\
\hline\hline
$dn/dN_x$ & $N_x=0$ & $N_x=1$ & $N_x=2$ & $N_x=3$ & $N_x=4$ & $N_x=5$ & $N_x=6$\\
\hline
$dn/dN_F$     &  0 (0)     & 1 (120)  & 2.5 (244) & 3 (225)  & 2.5 (115) & 1 (15) & 1 (1)\\
$dn/dN_{Z=1}$ &  4 (265)   & 2 (264)  & 2 (135)  & 1 (40)  & 1 (15)  & 0 (0) & 1 (1)\\
$dn/dN_{Z=2}$ &  5.5 (405) & 2.5 (195) & 1.5 (75) & 1.5 (45) & 0 (0)  & 0 (0) & 0 (0)\\
$dn/dN_{Z=3}$ &  8 (520)   & 2 (160)  & 1 (40)  & 0 (0)   & 0 (0)  & 0 (0) & 0 (0)\\
$dn/dN_{Z=4}$ & 10 (600)   & 1 (120)  & 0 (0)  & 0 (0)  & 0 (0)  & 0 (0) & 0 (0)\\
$dn/dN_{Z=5}$ & 10 (576)   & 1 (144)  & 0 (0)  & 0 (0)  & 0 (0)  & 0 (0) & 0 (0)\\
$dn/dN_{Z=6}$ & 10 (600)   & 1 (120)  & 0 (0)  & 0 (0)  & 0 (0)  & 0 (0) & 0 (0)\\
\hline
\end{tabular}%
\end{center}}
\end{table*}

\begin{table*}[htb!]
{\small Table 6: The multiplicity distribution $dn/dN_F$ of all
fragments and the multiplicity distribution $dn/dN_Z$ of the
fragments with charge $Z$ in excited $^{16}$O fragmentation in the
equal (unequal) probability partitioning method, where the
normalization is 22 (40320), which is the number of configurations
(exchanges).} \vspace{-10mm} {\small
\begin{center}
\begin{tabular}{cccccccccc}\\
\hline\hline
$dn/dN_x$ & $N_x=0$ & $N_x=1$ & $N_x=2$ & $N_x=3$ & $N_x=4$ & $N_x=5$ & $N_x=6$ & $N_x=7$ & $N_x=8$\\
\hline
$dn/dN_F$     &  0 (0)        & 1 (5040)    & 3.25 (12312) & 5 (12516)  & 5.25 (7301) & 3 (2660) & 2.5 (462) & 1 (28) & 1 (1)\\
$dn/dN_{Z=1}$ &  7 (14833)    & 4 (14832)   & 4 (7420)     & 2 (2464)   & 2 (630)     & 1 (112)  & 1 (28)    & 0 (0)  & 1 (1)\\
$dn/dN_{Z=2}$ &  9.25 (22449) & 5.5 (11340) & 4 (4494)     & 1.5 (1260) & 1.75 (777)  & 0 (0)    & 0 (0)     & 0 (0)  & 0 (0)\\
$dn/dN_{Z=3}$ & 15 (29120)    & 5 (8960)    & 2 (2240)     & 0 (0)      & 0 (0)       & 0 (0)    & 0 (0)     & 0 (0)  & 0 (0)\\
$dn/dN_{Z=4}$ & 19.25 (34272) & 2.5 (5544)  & 0.25 (504)   & 0 (0)      & 0 (0)       & 0 (0)    & 0 (0)     & 0 (0)  & 0 (0)\\
$dn/dN_{Z=5}$ & 19 (32256)    & 3 (8064)    & 0 (0)        & 0 (0)      & 0 (0)       & 0 (0)    & 0 (0)     & 0 (0)  & 0 (0)\\
$dn/dN_{Z=6}$ & 20 (33600)    & 2 (6720)    & 0 (0)        & 0 (0)      & 0 (0)       & 0 (0)    & 0 (0)     & 0 (0)  & 0 (0)\\
$dn/dN_{Z=7}$ & 21 (34560)    & 1 (5760)    & 0 (0)        & 0 (0)      & 0 (0)       & 0 (0)    & 0 (0)     & 0 (0)  & 0 (0)\\
$dn/dN_{Z=8}$ & 21 (35280)    & 1 (5040)    & 0 (0)        & 0 (0)      & 0 (0)       & 0 (0)    & 0 (0)     & 0 (0)  & 0 (0)\\
\hline
\end{tabular}%
\end{center}}
\end{table*}

\begin{figure*}[htb!]
\begin{center}
\includegraphics[width=15.5cm]{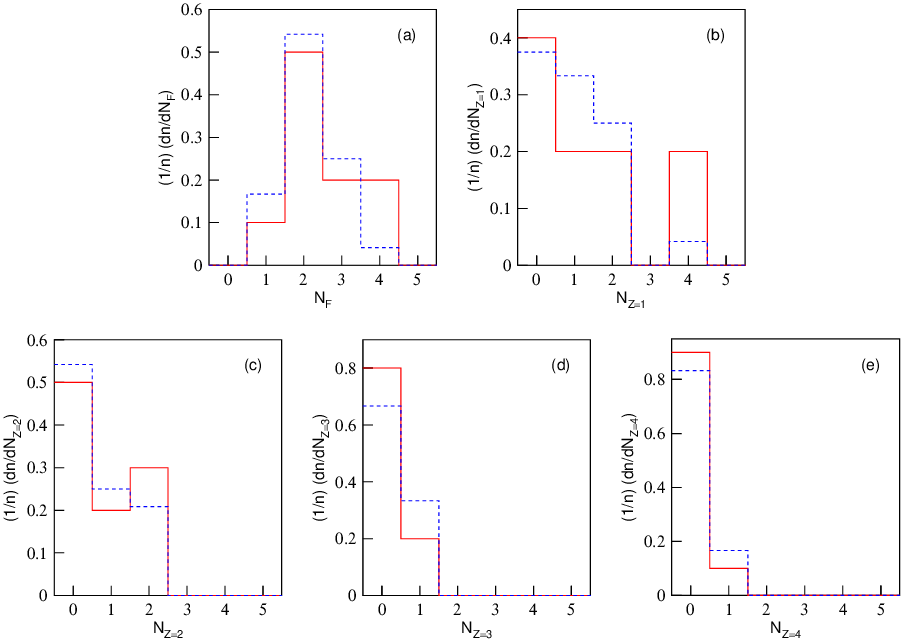}
\end{center}
\vskip-0.35cm {\small Figure 1: Multiplicity distributions of
nuclear fragments with different charges in $^{9}$Be
fragmentation. The solid (dashed) histograms represent the results
from the equal (unequal) probability partitioning method. Panel
(a) is for all fragments. Panels (b)--(e) are for the fragments
with charge $Z=1$, 2, 3, and 4, respectively.}
\end{figure*}

\begin{figure*}[htb!]
\begin{center}
\includegraphics[width=15.5cm]{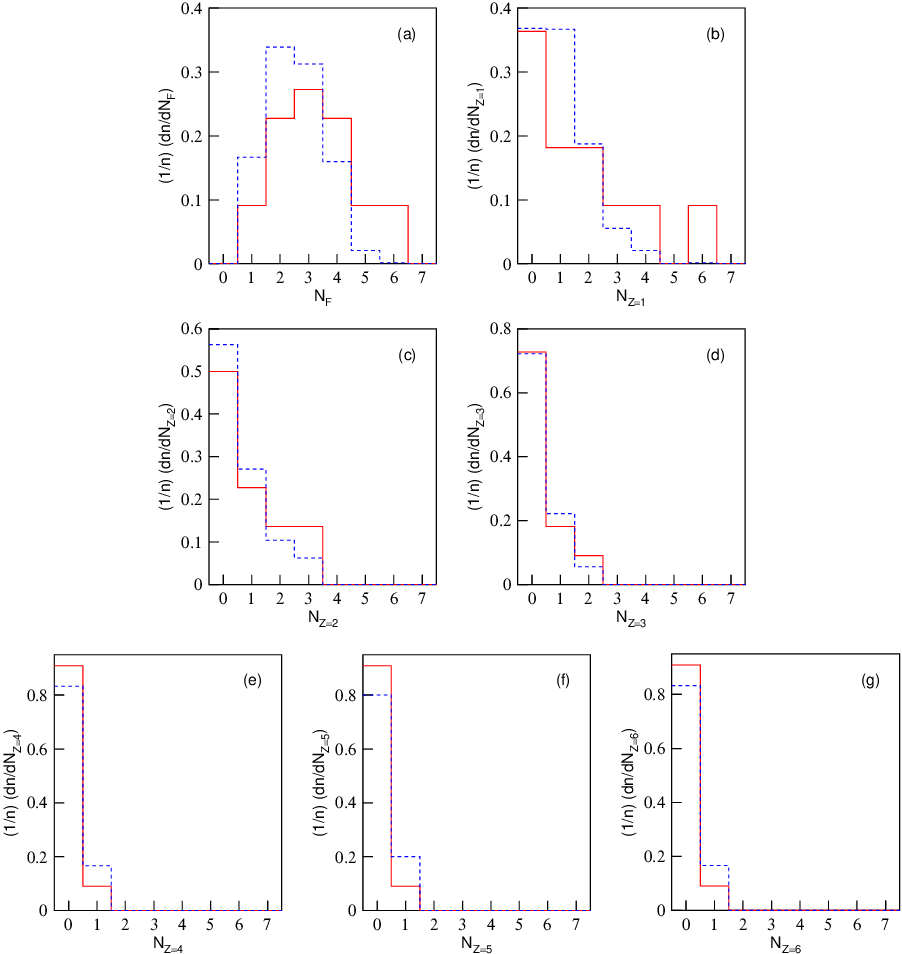}
\end{center}
\vskip-0.35cm {\small Figure 2: Multiplicity distributions of
nuclear fragments with different charges in $^{12}$C
fragmentation. The solid (dashed) histograms represent the results
from the equal (unequal) probability partitioning method. Panel
(a) is for all fragments. Panels (b)--(g) are for the fragments
with charge $Z=1$, 2, ..., and 6, respectively.}
\end{figure*}

\begin{figure*}[htb!]
\begin{center}
\includegraphics[width=15.5cm]{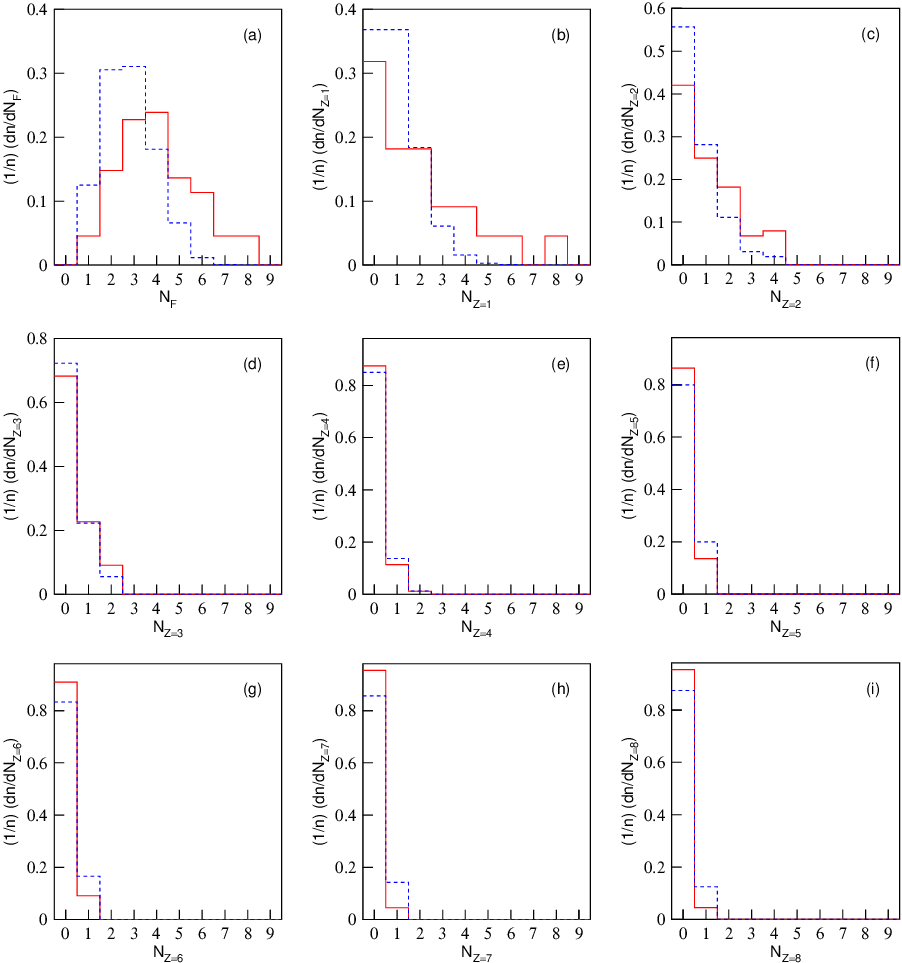}
\end{center}
\vskip-0.35cm {\small Figure 3: Multiplicity distributions of
nuclear fragments with different charges in $^{16}$O
fragmentation. The solid (dashed) histograms represent the results
from the equal (unequal) probability partitioning method. Panel
(a) is for all fragments. Panels (b)--(i) are for the fragments
with charge $Z=1$, 2, ..., and 8, respectively.}
\end{figure*}

{\section{Multiplicity distributions of nuclear fragments}}

After systematically deriving distinct
fragmentation channels using established partitioning methods, we
conduct a fragment-specific charge analysis within each channel.
This involves isolating individual fragments to precisely measure
and validate their charge states, leaving no detail unexamined.
Simultaneously, we meticulously track two key metrics: the total
fragment multiplicity per channel and the specific multiplicities
of fragments with predefined target charges. By applying the
corresponding weights from the partitioning methods to the
multiplicity data, we accurately determine the final multiplicity
distributions for both the complete fragment set and
charge-specified subsets, thereby revealing the inherent
probabilistic nature of nuclear fragmentation processes.

The $N_F$ distribution, $dn/dN_F$, and the $N_Z$ distribution,
$dn/dN_Z$, in fragmentation of excited $^9$Be, $^{12}$C, and
$^{16}$O nuclei are given in Tables 4, 5, and 6, respectively,
where $n$ denotes the frequency of $N_F$ occurring. The
multiplicity distributions $dn/dN_F$ and $dn/dN_Z$ are also the
yield distributions of nuclear fragments. The normalization
constants of $dn/dN_F$ and $dn/dN_Z$ in the partitioning methods
are the numbers of configurations (exchanges).

The normalized multiplicity distributions, $(1/n)(dn/dN_F)$ or
$(1/n)(dn/dN_Z)$, in excited $^9$Be, $^{12}$C, and $^{16}$O
fragmentation are displayed in Figures 1, 2, and 3, respectively.
The solid (dashed) histograms represent the results from the equal
(unequal) probability partitioning method. Figures 1(a), 2(a), and
3(a) are for the multiplicity distributions of all fragments. The
multiplicity distributions of the fragments with different $Z$ are
shown in different panels, where $Z=1$--4 in Figures 1(b)--1(e),
$Z=1$--6 in Figures 2(b)--2(g), and $Z=1$--8 in Figures 3(b)--3(i)
are for excited $^9$Be, $^{12}$C, and $^{16}$O fragmentation,
respectively. It is worth noting that the predicted frequency
distributions in Figures 1--3 are precise numerical values
(fractions) in each bin under a given scenario.

One can see from Tables 4--6 and Figures 1--3 that the
multiplicity distributions of the fragments with given $Z$ from
both the equal and unequal probability partitioning methods have a
quick decreasing trend in most cases. The larger the $Z$, the
closer the trends of the two results are. In the equal probability
partitioning method, the priority of 2He channel in $^9$Be
fragmentation is significant, and the priority of 3He (4He)
channel in $^{12}$C ($^{16}$O) fragmentation is not significant.
In the unequal probability partitioning method, the three cases do
not show an obvious priority. Figures 1(a), 2(a), and 3(a)
demonstrate peaks around the intermediate multiplicity, which are
naturally different from the multiplicity distribution of the
fragments with given $Z$.

In addition, the multiplicity distribution of the fragments with
$Z=2$ can be seen clearly. In particular, in the equal probability
partitioning method, $(dn/dN_{Z=2})/5=1.5/5=30\%$ for 2He channel
in excited $^9$Be fragmentation, $(dn/dN_{Z=2})/11=1.5/11=13.64\%$
for 3He channel in excited $^{12}$C fragmentation, and
$(dn/dN_{Z=2})/22=1.75/22=7.95\%$ for 4He channel in excited
$^{16}$O fragmentation. In the unequal probability partitioning
method, the three values are $5/24\approx20.8\%$, $45/720=6.25\%$,
and $777/40320\approx1.93\%$.

One can see that the difference between the two percentages from
the equal and unequal probability partitioning methods in $^9$Be
fragmentation is not too large, and that in $^{16}$O fragmentation
is quite large. If the $\alpha$-cluster structure does exist in
excited nuclei formed in $eA$ collisions at the EIC, one should
observe much more multi-He configuration than these percentages
(probabilities), which can be obtained from Tables 4--6 (Figures
1--3). Although there are some experimental reports on the
$\alpha$-cluster structure of excited
nucleus~\cite{c,19,20,21,22}, the related percentage or fraction
of multi-He configurations is significantly smaller than that
obtained through partitioning methods due to events with
multi-particle production included in the data
sample~\cite{19,20,21,22}. Thus, this fraction cannot be directly
compared with partitioning results.

An experimental study using nuclear emulsion~\cite{30} found that
the H+2He channel fraction in $^{10}$B fragmentation at a beam
energy of $E_{\rm beam}=1$ GeV/nucleon is 78\%. Based on this
finding, we estimate that the 2He channel fraction in $^{9}$Be
fragmentation at $E_{\rm beam}=1$ GeV/nucleon is approximately
78\%, possibly slightly higher due to fewer fragmentation channels
for $^{9}$Be compared to $^{10}$B. The inferred 2He channel
fraction in excited $^{9}$Be fragmentation is estimated to be
2.6--3.8 times that from partitioning methods. As a
non-conservative estimation, we set our judgment line for $\alpha$
clustering cases at twice the baseline percentages (probabilities)
without $\alpha$ clustering.

In this context, we assume that the experimental percentage of
multi-He events follows a Gaussian distribution with standard
deviation $\sigma$, predominantly concentrated within the range
[$0,4\sigma$] and centered around an expected value of $2\sigma$,
which serves as our baseline. If the experimental percentage
exceeds $4\sigma$, defined as twice the baseline and serving as
our threshold for judgment, one can draw conclusions regarding the
existence of $^3$He or $\alpha$ clustering with over 95\%
confidence level. According to this line, we conclude that $^3$He
or $\alpha$ clustering exists in excited $^{9}$Be formed during
peripheral collisions between $^{10}$B and nuclear emulsion at
$E_{\rm beam}=1$ GeV/nucleon.

Experimental data on the fragmentation of $^{9}$C, $^{10}$C, and
$^{11}$C at an energy of $E_{\rm beam}=1.2$ GeV/nucleon within
nuclear emulsion show fractions for the 3He channel as follows:
15.2\%, 5.3\%, and 17.5\%, respectively~\cite{31,32,33}.
Additionally, experiments on $^{16}$O fragmentation at $E_{\rm
beam}=3.65$ and 200 GeV/nucleon within nuclear emulsion indicate
fractions for the 4He channel as 12.5\% and 2.3\%
respectively~\cite{34}. The fraction of multi-He channels in the
fragmentation of excited $^{9,11}$C ($^{16}$O) formed at $E_{\rm
beam}=1.2$ (3.65) GeV/nucleon is more than double that predicted
by unequal probability partitioning, suggesting the presence of
$^3$He or $\alpha$ clustering in these excited nuclei. However,
for excited $^{10}$C formed at $E_{\rm beam}=1.2$ GeV/nucleon and
excited $^{16}$O formed at 200 GeV/nucleon, the multi-He channel
fractions do not exceed twice those from unequal probability
partitioning, indicating a stochastic result rather than $^3$He or
$\alpha$ clustering.

It should be noted that the errors in experimental data quoted
here are not available in refs.~\cite{30,31,32,33,34}. According
to the errors in data for other channels~\cite{30}, the relative
errors for the quoted data are estimated by us to be 15--21\%.
Generally, at $E_{\rm beam}=1.2$ GeV/nucleon, the fraction of the
3He channel in excited $^{10}$C fragmentation is significantly
lower than that in excited $^{9,11}$C due to its even-even nature,
which enhances its stability and reduces $^3$He or $\alpha$
clustering probabilities while increasing other fragmentation
channels' likelihoods. Additionally, original $\alpha$ clustering
presented in excited $^{16}$O at 3.65 GeV/nucleon is disrupted by
violent collisions at 200 GeV/nucleon. These collisions lead to
multi-particle production and participant nucleons separating from
$^{16}$O, making conditions for forming 4He clusters less
favorable. Furthermore, higher excitation energy achieved with
increased beam energy likely surpasses threshold energies needed
for forming such clusters; thus higher-excited $^{16}$O fragments
into multiple nucleons instead of favoring a 4He channel.

Based on the judgment line, the fractions of 2He (3He or 4He)
channel in excited $^9$Be ($^{12}$C or $^{16}$O) fragmentation
should be higher than 60\% (27.28\% or 15.9\%) if the equal
probability partitioning method is considered, or, 41.6\% (12.5\%
or 3.86\%) if the unequal probability partitioning method is
considered. Here, these percentages are obtained from twice the
values shown in Figures 1(c), 2(c), and 3(c), respectively,
according to the assumption of twice the baselines. One may note
that excited $^9$Be shows a significant 2He frequency and excited
$^{12}$C ($^{16}$O) does not show obvious enhancement of 3He
(4He). The reason is that $^9$Be has very few fragmentation
channels totally, and $^{12}$C ($^{16}$O) has relatively more
fragmentation channels totally. The tables and figures presented
in the present work can be regarded as a benchmark reference
result in which the $\alpha$-cluster model does not enter. We look
forward to the results of excited nuclear fragmentation at the
forthcoming EIC experiments to study the fraction of multi-He
configuration.

In addition, in the excited nucleus formed in $eA$ collisions, a
liquid-gas phase transition may also occur. In the above
discussions on nuclear fragmentation, the liquid-gas phase
transition is not taken into account in the calculations. If the
liquid-gas phase transition occurs, more light fragments should be
produced, causing the distribution of light fragment multiplicity
to deviate from the histogram in Figures 1--3, reducing the
probability of low multiplicity events and increasing the
probability of high multiplicity events. Meanwhile, heavy
fragments should not be produced, or their yield should be very
low. The results of this work can also provide reference for
whether liquid-gas phase transition occurs in the excited nucleus
in $eA$ collisions at the future EIC.

In $eA$ collisions, if the incident nucleus $A$ is very large, the
liquid-gas phase transition can occur in a part of the excited
nucleus. For the local area, where the phase transition has
occurred, many light fragments are expected to be emitted, and
there is no intermediate and heavy fragment emitted with them. For
the remainder area, where the phase transition has not happened,
the fragmentation is not special, in which the multiplicity
distribution of nuclear fragments should generally obey the
partitioning methods.

To ensure an accurate description as possible, the heaviest
fragments produced in an event---considered remnants of excited
nuclear fragmentation---should be excluded from analysis. For
genuine evaporation products, it must be acknowledged that they
originate from the fragmentation process involving a smaller
excited nucleus; thus, the partitioning methods should be
reapplied specifically for this smaller nucleus. Whether a
liquid-gas phase transition occurs in the overall or local area,
the proportion of light fragments with $Z=1$ should exceed twice
the baseline values when applying Gaussian distribution to the
considered probabilities. Furthermore, if experimental
measurements fall within the theoretical uncertainty range,
fragmentation may be interpreted as a consequence of a general
stochastic process.

Considering that excited nuclei undergo
liquid-gas phase transitions locally or globally, we take the
probability of the fragmentation channels including 2H--4H (2H or
4H) in $^9$Be, 3H--6H (3H, 4H, or 6H) in $^{12}$C, and 4H--8H (4H,
5H, 6H, or 8H) in $^{16}$O, from Tables 1--3, as the baseline
values. Based on the baseline values, the fractions of channels
including 2H--4H (3H--6H or 4H--8H) in excited $^9$Be ($^{12}$C or
$^{16}$O) fragmentation should be higher than 80\% (54.55\% or
45.45\%) if the equal probability partitioning method is
considered, or 58.33\% (15.56\% or 3.82\%) if the unequal
probability partitioning method is considered. Here, these
percentages are derived by doubling the baseline values. The
baseline values are assumed to be primarily concentrated within
the range $[0,4\sigma]$ when a Gaussian probability distribution
with width $\sigma$ is applied.

Beyond $\alpha$ clustering and liquid-gas phase
transitions---which may lead to significant deviations between
experimental multiplicity distributions and theoretical
models---other nuclear effects exert only minor influences on
experimental outcomes. These nuclear effects involved include
non-uniform nucleon number density distributions (the neutron skin
structure of heavy nuclei), symmetrical energy characteristics of
nuclear matter, two- or multi-nucleon correlations within nuclei,
as well as stopping power or transparency phenomena associated
with nuclear interactions. Here by the slight effects it is meant
that both the effects themselves and their impact can be neglected
in studying the multiplicity distribution of nuclear fragments.

The reason why other nuclear effects are small is that they mainly
affect the momentum distribution of nucleons inside the nucleus.
Due to limited strength, the other nuclear effects mentioned above
are not sufficient to affect the formation of nuclear fragments
with given charge $Z$, though they affect the neutron numbers in
emitted isotopes. As a result, they also affect the kinetic energy
and emitting direction of nuclear fragments. In short, the
transverse momenta and polar angles of nuclear
fragments are significantly affected, while the charges and
multiplicity of nuclear fragments are slightly affected.
\\

\begin{figure*}[htb!]
\begin{center}
\includegraphics[width=11.5cm]{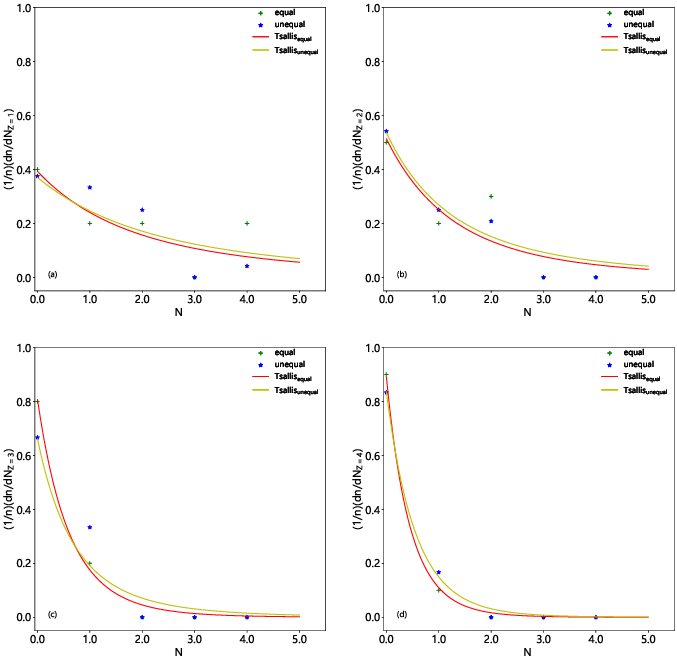}
\end{center}
\vskip-0.35cm {\small Figure 4: Multiplicity distributions of
nuclear fragments with $Z=1$ (a), 2 (b), 3 (c), and 4 (d) in
$^{9}$Be fragmentation. The crosses (asterisks) represent the
results from the equal (unequal) probability partitioning method,
which are cited from the solid (dashed) histograms in Figure 1.
The corresponding results fitted by the Tsallis probability
density function [Eq. (7)] are presented by the red (yellow)
curves.}
\end{figure*}

\begin{figure*}[htb!]
\begin{center}
\includegraphics[width=11.5cm]{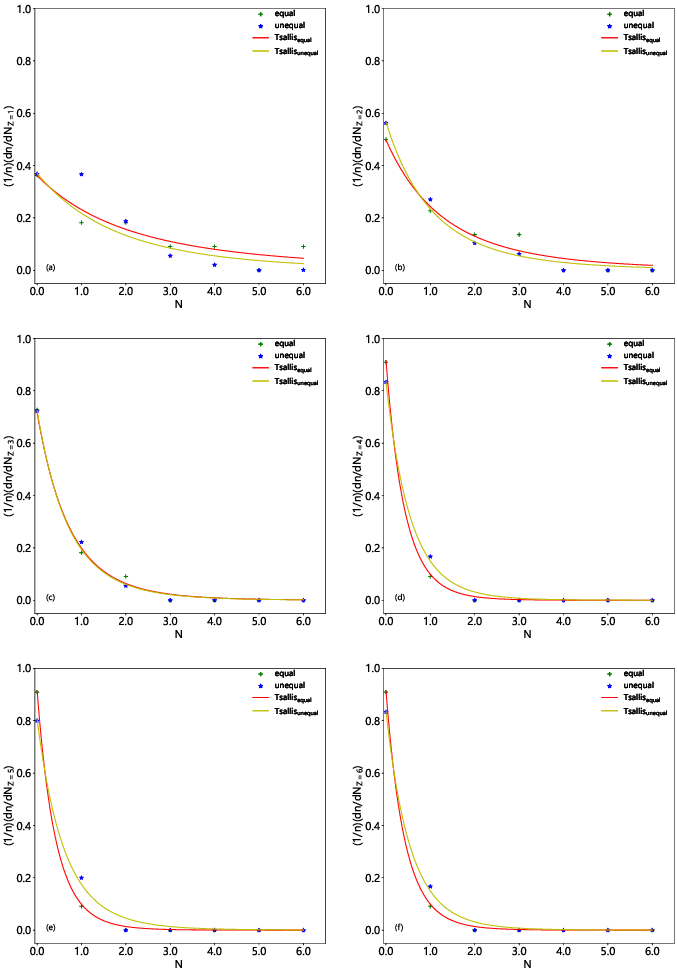}
\end{center}
\vskip-0.35cm {\small Figure 5: Multiplicity distributions of
nuclear fragments with $Z=1$ (a), 2 (b), 3 (c), 4 (d), 5 (e), and
6 (f) in $^{12}$C fragmentation. The crosses (asterisks) represent
the results from the equal (unequal) probability partitioning
method, which are cited from the solid (dashed) histograms in
Figure 2. The corresponding results fitted by the Tsallis
probability density function [Eq. (7)] are presented by the red
(yellow) curves.}
\end{figure*}

\begin{figure*}[htb!]
\begin{center}
\vskip-1.02cm
\includegraphics[width=11.5cm]{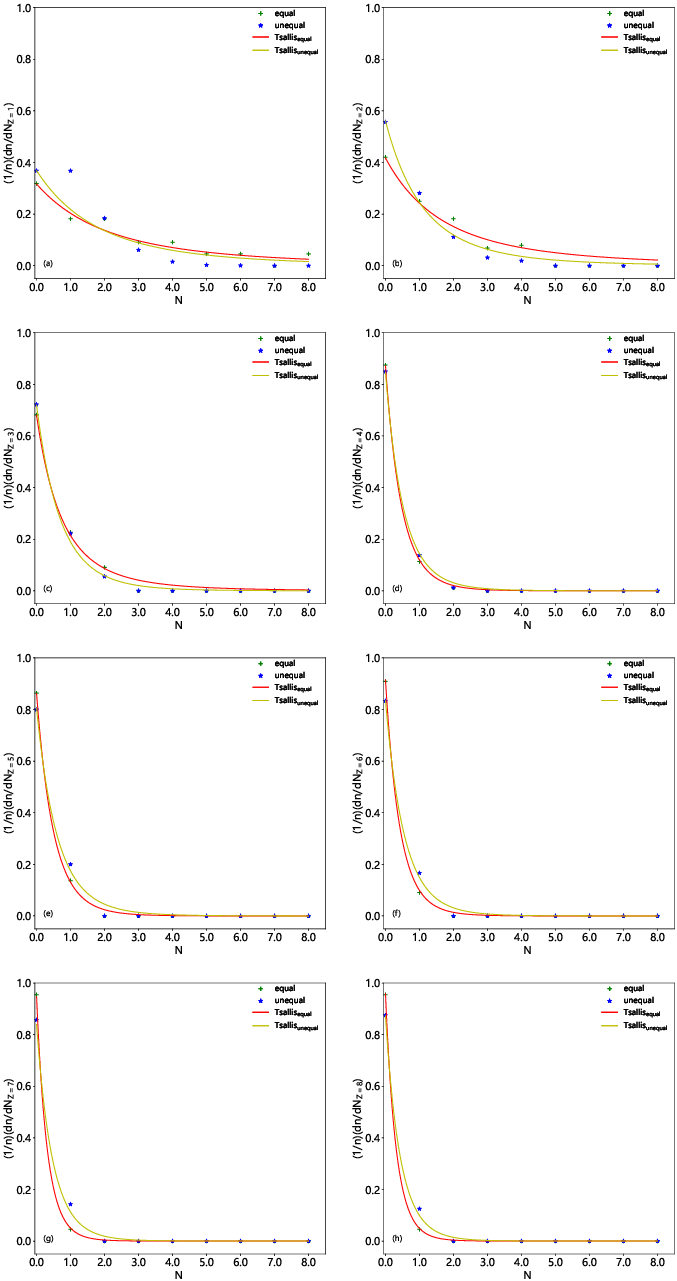}
\end{center}
\vskip-0.35cm {\small Figure 6: Multiplicity distributions of
nuclear fragments with $Z=1$ (a), 2 (b), 3 (c), 4 (d), 5 (e), 6
(f), 7 (g), and 8 (h) in $^{16}$O fragmentation. The crosses
(asterisks) represent the results from the equal (unequal)
probability partitioning method, which are cited from the solid
(dashed) histograms in Figure 3. The corresponding results fitted
by the Tsallis probability density function [Eq. (7)] are
presented by the red (yellow) curves.}
\end{figure*}

\begin{figure*}[htb!]
\begin{center}
\includegraphics[width=17.0cm]{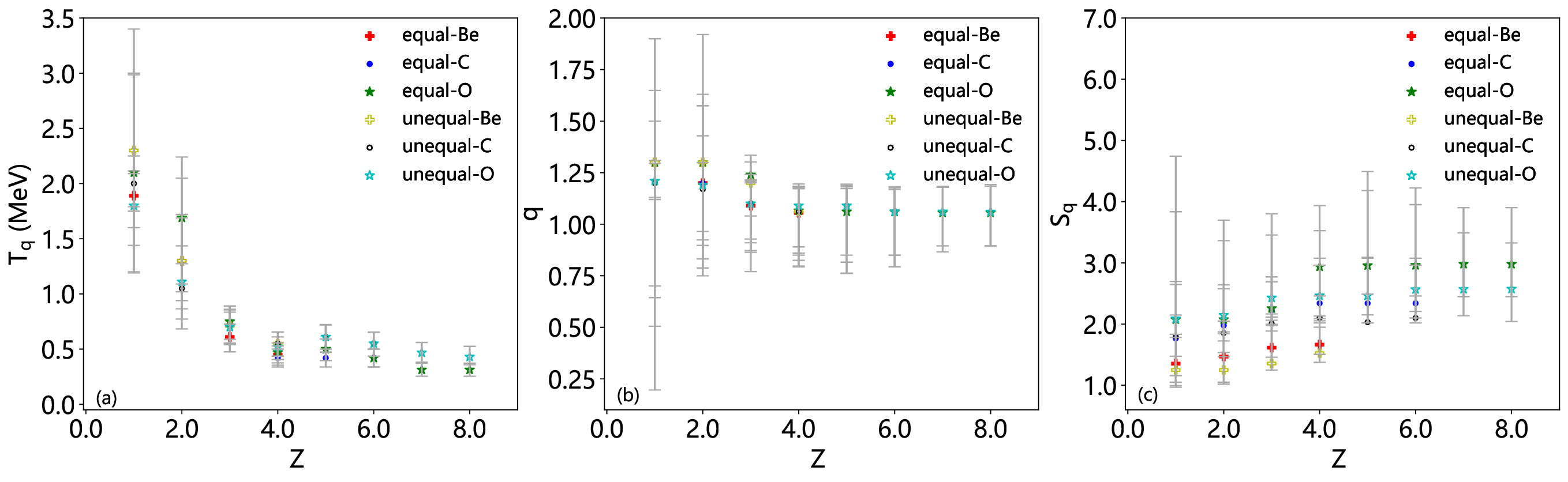}
\end{center}
\vskip-0.35cm {\small Figure 7: Dependencies of the Tsallis
nonextensive parameters---(a) generalized temperature $T_q$, (b)
entropy index $q$, and (c) $q$-entropy $S_q$---on fragment charge
number $Z$ in nuclear fragmentation reactions of $^{9}$Be
(crosses), $^{12}$C (circles), and $^{16}$O (asterisks),
determined via equal-probability (filled symbols) and
unequal-probability (open symbols) partitioning methods. }
\end{figure*}

{\section{Nonextensive parameters from
multiplicity distributions of nuclear fragments}}

Tsallis statistics represents an extension and generalization of
Boltzmann-Gibbs statistics. It introduces the nonextensive entropy
index $q$, built upon the foundational concept of Boltzmann-Gibbs
statistics---the Boltzmann-Gibbs entropy---to construct a new form
of entropy, known as the Tsallis (nonextensive or non-additive)
entropy $S_q$~\cite{55,56,57,58,59}. The Tsallis nonextensive
statistical framework extends traditional Boltzmann-Gibbs
statistics to describe systems characterized by long-range
interactions, non-equilibrium dynamics, or fractal-like
structures. In the limit $q\rightarrow1$, both the entropy $S_q$
and the distribution function of Tsallis statistics reduce to
their Boltzmann-Gibbs counterparts. Thus, Boltzmann-Gibbs
statistics can be regarded as a special case of Tsallis statistics
when $q=1$.

The probability density function used in Tsallis statistics has
different forms or revisions~\cite{55,56,57,58,59}. For the
multiplicity $N$ distribution of nuclear fragments, we use
\begin{align}
P(N) =P(0) \left[1-\frac{(1-q)N}{T_q}\right]^{\frac{1}{1-q}},
\end{align}
where $P(0)$ denotes the probability of events with zero
multiplicity for fragments of a given charge number, $T_q$ is a
generalized temperature which measures the average energy per
degree of freedom in the generalized equilibrium state, distinct
from the conventional temperature in thermal equilibrium systems,
and $q$ denotes entropy index which quantifies the degree of
non-extensivity of the system. A value of $q=1$ corresponds to the
Boltzmann-Gibbs limit, while $q>1$ indicates enhanced nonextensive
behavior due to strong correlations or non-equilibrium effects.

In the description for multiplicity distribution of nuclear
fragments, the form of nonextensive or non-additive $q$-entropy
$S_q$ is written by~\cite{55,56,57,58,59}
\begin{align}
S_q =k \frac{1-\sum_N P_N^q}{q-1},
\end{align}
where $k$ is the Boltzmann constant which equals to 1 in the
natural units, $P_N$ [$=P(N)$] denotes the probability of the
nuclear fragment with multiplicity $N$ and satisfies to $\sum_N
P_N=1$. $S_q$ measures the degree of disorder or complexity in the
system, accounting for non-additive contributions from correlated
subsystems.

We present the multiplicity distributions of nuclear fragments
with $Z=1$--4 (a--d), $Z=1$--6 (a--f), and $Z=1$--8 (a--h) in
$^{9}$Be, $^{12}$C, and $^{16}$O fragmentations in Figures 4--6,
respectively. The crosses (asterisks) represent the results from
the equal (unequal) probability partitioning method, which are
cited from the solid (dashed) histograms in Figures 1--3,
respectively. The corresponding results fitted by the Tsallis
probability density function [Eq. (7)] are presented by the red
(yellow) curves. When judging the goodness of fit, the coefficient
of determination $R^2=1-{\rm (RSS/TSS)}$ is used,
where ${\rm RSS}=\sum (y_i-\hat{y}_i)^2$ represents the Residual
Sum of Squares, ${\rm TSS}=\sum (y_i-\bar{y})^2$ denotes the Total
Sum of Squares, $y_i$ is the actual observed value of the $i$-th
data point, $\hat{y}_i$ is the model-predicted value corresponding
to the $i$-th data point, and $\bar{y}$ is the mean of the
observed values. The values of $T_q$, $q$, and $R^2$ derived from
Figures 4--6 are summarized in Table 7. The closer $R^2$ is to 1,
the better the model's fit. One can see that in
most cases, the Tsallis probability density function can
approximately fit the multiplicity distribution of nuclear
fragments with different charge numbers.

\begin{table*}[htb!]
{\small Table 7: Values of $T_q$, $q$, and $R^2$ used for the curves in Figures 4--6 under
equal and unequal probability partitioning methods. Due to retaining three decimal places and rounding the fourth decimal place, some $R^2$ values are 1.000.\\} \vspace{-8mm} {\small
\begin{center}
\begin{tabular}{cccc|ccc}\\
\hline\hline
Figure &             & Equal &    &  & Unequal &  \\
\cline{2-7}
number & $T_q$ (MeV) & $q$ & $R^2$  & $T_q$ (MeV) & $q$ & $R^2$ \\
\hline
4(a)   & $1.890^{+0.225}_{-0.450}$ & $1.300^{+0.200}_{-1.104}$ & 0.620 & $2.300^{+0.685}_{-0.550}$ & $1.300^{+0.349}_{-0.265}$ & 0.728 \\
4(b)   & $1.300^{+0.750}_{-0.210}$ & $1.200^{+0.430}_{-0.276}$ & 0.787 & $1.300^{+0.389}_{-0.435}$ & $1.300^{+0.275}_{-0.460}$ & 0.920 \\
4(c)   & $0.610^{+0.116}_{-0.135}$ & $1.090^{+0.119}_{-0.320}$ & 0.994 & $0.710^{+0.151}_{-0.170}$ & $1.200^{+0.102}_{-0.550}$ & 0.926 \\ 
4(d)   & $0.450^{+0.078}_{-0.096}$ & $1.055^{+0.117}_{-0.230}$ & 0.999 & $0.550^{+0.105}_{-0.550}$ & $1.062^{+0.122}_{-0.795}$ & 0.997 \\
\hline
5(a)   & $2.300^{+1.101}_{-0.510}$ & $1.300^{+0.600}_{-0.170}$ & 0.885 & $1.810^{+0.250}_{-0.810}$ & $1.155^{+0.095}_{-0.556}$ & 0.825 \\
5(b)   & $1.300^{+0.420}_{-0.281}$ & $1.200^{+0.229}_{-0.235}$ & 0.963 & $1.050^{+0.224}_{-0.278}$ & $1.170^{+0.129}_{-0.273}$ & 0.990 \\
5(c)   & $0.730^{+0.161}_{-0.141}$ & $1.100^{+0.141}_{-0.172}$ & 0.996 & $0.710^{+0.145}_{-0.155}$ & $1.100^{+0.115}_{-0.236}$ & 0.997 \\
5(d)   & $0.420^{+0.080}_{-0.082}$ & $1.060^{+0.122}_{-0.210}$ & 1.000 & $0.550^{+0.105}_{-0.115}$ & $1.062^{+0.115}_{-0.269}$ & 0.998 \\
5(e)   & $0.420^{+0.080}_{-0.082}$ & $1.060^{+0.122}_{-0.210}$ & 1.000 & $0.610^{+0.112}_{-0.137}$ & $1.090^{+0.104}_{-0.328}$ & 0.994 \\
5(f)   & $0.420^{+0.080}_{-0.082}$ & $1.060^{+0.122}_{-0.210}$ & 1.000 & $0.550^{+0.105}_{-0.115}$ & $1.062^{+0.115}_{-0.269}$ & 0.998 \\
\hline
6(a)   & $2.100^{+0.090}_{-0.500}$ & $1.300^{+0.600}_{-0.180}$ & 0.944 & $1.500^{+0.290}_{-0.599}$ & $1.200^{+0.100}_{-0.509}$ & 0.846 \\
6(b)   & $1.690^{+0.550}_{-0.750}$ & $1.300^{+0.275}_{-0.512}$ & 0.958 & $1.110^{+0.325}_{-0.427}$ & $1.190^{+0.173}_{-0.358}$ & 0.988 \\
6(c)   & $0.750^{+0.137}_{-0.202}$ & $1.240^{+0.095}_{-0.330}$ & 0.995 & $0.700^{+0.135}_{-0.153}$ & $1.100^{+0.101}_{-0.228}$ & 0.997 \\
6(d)   & $0.470^{+0.087}_{-0.095}$ & $1.066^{+0.116}_{-0.176}$ & 1.000 & $0.520^{+0.090}_{-0.115}$ & $1.090^{+0.105}_{-0.230}$ & 0.999 \\
6(e)   & $0.500^{+0.092}_{-0.103}$ & $1.061^{+0.112}_{-0.245}$ & 0.999 & $0.610^{+0.109}_{-0.138}$ & $1.090^{+0.098}_{-0.328}$ & 0.995 \\
6(f)   & $0.420^{+0.080}_{-0.082}$ & $1.060^{+0.121}_{-0.210}$ & 1.000 & $0.550^{+0.102}_{-0.115}$ & $1.062^{+0.107}_{-0.268}$ & 0.998 \\
6(g)   & $0.313^{+0.060}_{-0.059}$ & $1.056^{+0.128}_{-0.161}$ & 1.000 & $0.468^{+0.092}_{-0.087}$ & $1.061^{+0.118}_{-0.195}$ & 0.998 \\
6(h)   & $0.313^{+0.060}_{-0.059}$ & $1.056^{+0.128}_{-0.161}$ & 1.000 & $0.430^{+0.095}_{-0.072}$ & $1.060^{+0.132}_{-0.165}$ & 0.999 \\
\hline
\end{tabular}%
\end{center}}
\end{table*}

The dependencies of Tsallis nonextensive parameters---including
(a) $T_q$, (b) $q$, and (c) $S_q$---on the fragment charge number
$Z$ for nuclear fragmentation reactions of $^{9}$Be (crosses),
$^{12}$C (circles), and $^{16}$O (asterisks) are shown in Figure
7, where $T_q$ and $q$ listed in Table 7 are
extracted from the fit of multiplicity distribution via Eq. (7)
and $S_q$ is obtained due to Eq. (8). These results are derived
using two distinct partitioning approaches: the equal-probability
method (represented by filled symbols) and the unequal-probability
method (represented by open symbols). The error
bars in the free parameter figures are obtained by the $\chi^2$ 
profile method with a 95\% confidence level. One can see the 
tendencies of the considered nonextensive parameters.

The generalized temperature $T_q$ exhibits a decreasing trend with
increasing fragment charge number $Z$ for all three excited nuclei
($^{9}$Be, $^{12}$C, and $^{16}$O). This observation suggests that
heavier fragments, which carry a larger proportion of the parent
nucleus's charge, are associated with lower effective
temperatures. Physically, this can be interpreted as a result of
the more ordered internal structure and lower excitation energy of
heavier fragments, as they tend to retain more of the parent
nucleus's initial stability. Furthermore, a systematic
independence on the excited nucleus mass is observed: the $T_q$
values ($\sim0.5$--2 MeV) from multiplicity distribution of
$^{16}$O fragments are consistently in agreement with those from
multiplicity distributions of $^{12}$C and $^{9}$Be fragments at
the same $Z$ due to all the three nuclei are light nucleus. It is
expected that heavier excited nuclei produce fragments with more
stable configurations and lower average excitation energies,
likely due to their higher binding energy per nucleon and more
favorable fragmentation pathways that minimize the release of
excess energy. Notably, the difference between the
equal-probability and unequal-probability partitioning methods is
not significant in the whole $Z$ region.

The values ($\sim1.05$--1.3) of entropy index $q$ are found to
remain nearly unchanged within the uncertainty range, with increasing $Z$ across the three excited nuclei.  
Similar $q$ values support the possibility that collective excitation and surface effects may play a dominant role in high-energy fragmentation of light nuclei, rather than bulk behavior. The values of $q$ deviate from 1 for all fragments, confirming the strong
nonextensive nature of nuclear fragmentation reactions. This
deviation from the Boltzmann-Gibbs limit ($q=1$) is attributed to
the long-range nucleon-nucleon interactions, non-equilibrium
dynamics during the fragmentation process, and the fractal-like
structure of the phase space accessible to the fragments. Considering the uncertainty range, $q$ has a probability of being smaller than 1, which indicates that the nuclear fragmentation system tends to suppress high-energy excited states, causing the particle distribution to be more concentrated in low-energy regions, which may occur in some confined or strongly dissipative systems, although less common in nuclear physics.

The $q$-entropy $S_q$ also remains almost within the uncertainty range, with the fragment
charge number $Z$ for all three excited nuclei. This indicates that the microstructural diversity or information uncertainty levels of these three types of light nucleus fragmentation final states are similar, meaning that the way and degree of system ``chaos" are statistically equivalent, despite differences in fragment $Z$ distribution. These three types of light nuclei exhibit statistical self-similarity during fragmentation, suggesting that one can use a unified and effective theoretical model (such as Tsallis statistics) to describe the dynamics of light nucleus fragmentation, without the need to fit parameters separately for each nucleus.
The difference between the two partitioning methods can be neglected for the $q$-entropy $S_q$, even at large $Z$ values. This
indicates that the overall degree of disorder in the system is not
too sensitive to the choice of partitioning method, as that for
the generalized temperature $T_q$ and the entropy index $q$.
However, the unequal-probability method generally yields slightly
smoother $S_q(Z)$ curves, as it accounts for the non-uniform
probabilities of different fragmentation channels.

As mentioned above, the equal-probability partitioning method
assumes that all possible fragmentation channels are equally
likely, providing a simplified approach to parameter extraction.
However, this assumption may not hold in reality, as certain
fragmentation pathways may be favored due to quantum mechanical
effects (e.g., shell structure) or energetic considerations (e.g.,
minimum energy configurations). In contrast, the
unequal-probability partitioning method incorporates dynamical
weights based on the physical likelihood of each fragmentation
channel, offering a more realistic description of the reaction
mechanism. As a result, this method generally leads to more
reasonable $T_q$ and $q$ values (when accounting for preferential
pathways). The discrepancy between the two methods serves as a
valuable indicator of the reliability of the extracted parameters.
In regions where the differences are significant, it highlights
the need for careful consideration of the underlying fragmentation
dynamics and the choice of statistical framework.

Our analysis of the dependencies of Tsallis nonextensive
parameters on the fragment charge number $Z$ provides valuable
insights into the statistical nature of nuclear fragmentation
reactions. The observed trends in $T_q$, $q$, and $S_q$ with $Z$
and excited nucleus mass show that nuclear fragmentation reactions
result in a generalized equilibrium state that deviates
significantly from the traditional thermal equilibrium, as
evidenced by the nonextensive parameter values ($q\neq1$). Heavier fragments exhibit lower effective temperatures of emission source,
but similar nonextensive behavior and similar nonextensive entropy
compared to lighter fragments. These similarities reflect the consistency in statistical behavior and dynamic evolution, and the universal mechanism of fragmentation process.

The choice of partitioning method (equal-probability vs.
unequal-probability) has no notable impact on the extracted
parameters, particularly for the generalized temperature $T_q$ and
the entropy index $q$. However, the unequal-probability method,
which accounts for the physical likelihood of different
fragmentation channels, should provide a more accurate and
self-consistent description of the reaction dynamics. Our findings
underscore the utility of the Tsallis nonextensive statistical
framework in characterizing complex nuclear reactions and
highlight the importance of considering non-equilibrium effects
and correlated dynamics in such systems.

Before the summary and conclusion section, we
would like to emphasize that while Tsallis statistics demonstrate
strong applicability in fitting fragment multiplicity
distributions, the underlying physical mechanisms merit further
exploration. A core feature of Tsallis statistics is its
nonextensivity, characterized by the parameter $q$, which allows
it to describe systems with long-range interactions or significant
fluctuations. During fragment production, the complexity of the
collision process and diversity of intermediate states give rise
to substantial system fluctuations that cannot be captured by
traditional equilibrium statistics. Tsallis nonextensive
statistics, however, can precisely account for distribution
deviations caused by such fluctuations, enabling successful
fitting of multiplicity distributions.

In addition, Tsallis statistics can effectively
describe particle production behavior during the formation of
Quark-Gluon Plasma (QGP). Fragment production and QGP formation
may share key similarities---both involve strong interactions and
particle cascade processes in non-equilibrium states, which could
be one potential reason for the applicability of Tsallis
statistics in nuclear fragmentation. Our research reveals that the
nearly invariant trend of parameter $q$ within the uncertainty range as $Z$ changes is closely linked to energy dissipation and particle correlation during fragmentation: as $Z$ increases, the degree to which $q$ deviates from 1 does not change, indicating similar local fluctuations within the system. This finding further validates the use of Tsallis
statistics for describing non-equilibrium fragment production
processes. 
\\

{\section{Summary and conclusion}}

Various configurations of nuclear fragments
resulting from the fragmentation of excited $^9$Be, $^{12}$C, and
$^{16}$O nuclei---expected to form in $eA$ collisions at the
EIC---are investigated using both equal and unequal probability
partitioning methods. The multiplicity distributions for all
fragments as well as those with charge $Z$ are derived. In
comparison to results obtained from these partitioning methods,
experiments suggest that multi-$\alpha$ configurations should
exhibit a significantly high probability according to the
$\alpha$-cluster model. We anticipate that the structure of
excited nuclei featuring an $\alpha$-cluster will be clearly
manifested and further validated in future studies.
According to our predictions, the fraction of the
2He (3He or 4He) channel in excited $^9$Be ($^{12}$C or $^{16}$O)
fragmentation should exceed 60\% (27.28\%, 15.9\%) under the equal
probability partitioning method, or 41.6\% (12.5\%, 3.86\%) under
the unequal probability partitioning method, with over 95\%
confidence level.

Additionally, findings from this work can serve as a reference for
assessing whether a liquid-gas phase transition occurs within
excited nucleus in $eA$ collisions. Should such a phase transition
take place experimentally, an increased observation of light
fragments is expected alongside minimal detection of heavy
fragments. For very heavy excited nuclei, it is plausible that
liquid-gas phase transition could occur in specific region where
numerous light fragments are evaporated while other area undergoes
fragmentation process or remaining smaller nucleus; this
fragmentation process may deviate from traditional partitioning
methods if significant $\alpha$-clustering is present.
We predict that the fraction of channels spanning
2H--4H (3H--6H or 4H--8H) in excited $^9$Be ($^{12}$C or $^{16}$O)
fragmentation should exceed 80\% (54.55\% or 45.45\%) under the
equal probability partitioning method, or 58.33\% (15.56\% or
3.82\%) under the unequal probability partitioning method, with
over 95\% confidence level.

In the framework of Tsallis statistics, the nonextensive
parameters $T_q$, $q$, and $S_q$ are obtained from the
multiplicity distribution of nuclear fragments with given $Z$.
With the increase of $Z$, $T_q$ decreases, while both $q$ and $S_q$ remain nearly unchanged within the uncertainty range. Our work shows that fragmentation of nuclear remnants
in electron-nucleus collisions at high energy is a nonextensive
process ($q\approx1.05$--1.3) with a temperature of
$T_q\approx0.5$--2 MeV. This work reveals the transformative
utility of the Tsallis nonextensive statistical framework in
decoding the previously uncharacterized complexities of nuclear
fragmentation reactions, challenging traditional equilibrium-based
models and emphasizing the urgent need to integrate
non-equilibrium effects and correlated dynamics into the core of
nuclear reaction theory.

Before the end of this paper, we would like to
point out that our study relies solely on the $\alpha$-cluster
model and two partitioning methods (equal and unequal
probability). However, the actual fragmentation process in $eA$
collisions likely involves more complex correlated dynamics and
non-equilibrium effects not captured by these models. The Tsallis
statistics, while revealing nonextensive features, remains a
phenomenological approach lacking a microscopic foundation for the
fragmentation mechanism. The predicted fragmentation channel
fractions (multi-He fragment dominance) and liquid-gas phase
transition signatures (multi-H fragment dominance) require future
experimental verification. Moreover, we have only examined three
not-too-heavy excited nuclei; the fragmentation behavior of
heavier nuclei (e.g., medium and heavy mass) under $eA$ collisions
remains unexplored, limiting the generalizability of our
conclusions.

Future work should develop a microscopic model
that incorporates $\alpha$-clustering and non-equilibrium dynamics
into the fragmentation process, going beyond phenomenological
Tsallis statistics. This could involve coupling the
$\alpha$-cluster model with dynamical equations to describe the
time evolution of excited nuclei. Modelers should collaborate with
experimental groups at the EIC to design dedicated measurements of
$\alpha$-cluster configurations and light fragment multiplicities,
and to develop advanced data analysis techniques for improved
detection efficiency and fragment identification. The theoretical
framework should also be extended to medium and heavy mass nuclei,
investigating the liquid-gas phase transition in these systems and
testing the universality of fragmentation patterns observed in
light nuclei. Finally, a microscopic derivation of the Tsallis
nonextensive statistical parameters would help establish a more
fundamental connection between the fragmentation process and the
underlying nuclear dynamics.
\\
\\
{\bf Author Contributions:} The authors contributed to the paper
in this way: Conceptualization, F.H.L. and Kh.K.O.; Methodology,
F.H.L. and Kh.K.O.; Software, T.T.D.; Validation, F.H.L. and
Kh.K.O.; Formal analysis, T.T.D.; Investigation, T.T.D., S.B. and
H.L.L.; Resources, F.H.L.; Data curation, T.T.D., S.B. and H.L.L.;
Writing---original draft preparation, T.T.D.; Writing---review and
editing, F.H.L. and Kh.K.O.; Visualization, T.T.D., S.B. and
H.L.L.; Supervision, F.H.L. and Kh.K.O.; Project administration,
F.H.L.; Funding acquisition, F.H.L. and Kh.K.O. All authors have
read and agreed to the published version of the manuscript.
\\
\\
{\bf Funding:} The work of Shanxi Group was supported by the
National Natural Science Foundation of China under Grant No.
12147215, the Shanxi Provincial Basic Research Program (Natural
Science Foundation) under Grant No. 202103021224036, and the Fund
for Shanxi ``1331 Project" Key Subjects Construction. The work of
K.K.O. was supported by the Agency of Innovative Development under
the Ministry of Higher Education, Science and Innovations of the
Republic of Uzbekistan within the fundamental project No.
F3-20200929146 on analysis of open data on heavy-ion collisions at
RHIC and LHC.
\\
\\
{\bf Institutional Review Board Statement:} Not applicable.
\\
\\
{\bf Informed Consent Statement:} Not applicable.
\\
\\
{\bf Data Availability Statement:} The data used to support the
findings of this study are included within the article and are
cited at relevant places within the text as references.
\\
\\
{\bf Conflicts of Interest:} The authors declare that there are no
conflicts of interest regarding the publication of this paper. The
funders had no role in the design of the study; in the collection,
analysis, or interpretation of the data; in the writing of the
manuscript; or in the decision to publish the results.
\\
\\
{\bf ORCID}
\\
Fu-Hu Liu https://orcid.org/0000-0002-2261-6899
\\
Khusniddin K. Olimov https://orcid.org/0000-0002-1879-8458
\\
\\

\end{document}